\documentclass[aps,prl,twocolumn,groupedaddress,showpacs]{revtex4}

\usepackage{graphicx}
\usepackage{dcolumn}
\usepackage{bm}
\usepackage{hyperref}
\usepackage[mathlines]{lineno}

\usepackage{amsmath}

\begin{document}




\title{Cluster dynamical mean field theory of quantum phases on a honeycomb lattice}




\author{Rong-Qiang He}
\author{Zhong-Yi Lu}
\affiliation{Department of Physics, Renmin University of China, Beijing 100872, China}


\date{\today}

\begin{abstract}

We have studied the ground state of the half-filled Hubbard model on a honeycomb lattice
by performing the cluster dynamical mean field theory calculations with exact
diagonalization on the cluster-impurity solver. Through using elaborate numerical
analytic continuation, we identify the existence of a `spin liquid' from the on-site
interaction $U=0$ to $U_c$ (between $4.6t$ and $4.85t$) with a smooth crossover
correspondingly from the charge fluctuation dominating phase into the charge correlation
dominating phase. The semi-metallic state exits only at $U=0$. We further find that the
magnetic phase transition at $U_c$ from the `spin liquid' to the N\'{e}el
antiferromagnetic Mott insulating phase is a first-order quantum phase transition. We
also show that the charge fluctuation plays a substantial role on keeping the `spin
liquid' phase against the emergence of a magnetic order.

\end{abstract}


\maketitle


Quantum phase transition is a fascinating physics subject, which describes an abrupt
change of the ground state of a quantum many-body system tuned by a non-thermal physical
parameter, often accompanied with a novel quantum emergence phenomenon \cite{book1}. As a
canonical quantum phase transition, the Mott transition, from metallic to insulating
state tuned by electronic Coulomb interaction, is one of the most celebrated and
difficult problems in condensed matter physics \cite{book2}. The resultant insulating
state, namely Mott insulator, usually adopts spontaneous symmetry breaking in two and
three spatial dimension to form a long-range antiferromagnetic (AFM) order to release the
spin entropy due to localized electrons. Theoretically, the simplest model to capture
such physics is the standard one-band half-filled Hubbard model. In the large Coulomb
interaction limit this model reduces to a standard Heisenberg model with an AFM order in
its ground state.

Nevertheless, an insulating ground state without any spontaneous symmetry breaking,
namely spin liquid, may arise if there is frustration \cite{anderson1,balents}. Actually
the spin liquid is a genuine Mott insulating state in a sense that it is adiabatically
separated from a band insulator. Spin liquid has been one of the most intriguing issues
in condensed matter physics since it was introduced nearly forty years ago
\cite{anderson1} and continuously in intense research since it was further proposed to be
a parent phase to likely lead to high T$_c$ superconductivity \cite{anderson2}. However,
a spin liquid had not been verified for a two-dimensional (2D) standard Hubbard or
Heisenberg model until a recent quantum Monte Carlo simulation was done for correlated
fermions on a honeycomb lattice \cite{Meng10}. Through finite size extrapolation the
simulation surprisingly shows that a spin liquid emerges between the semi-metallic and
AFM Mott insulating states. This has stimulated a new surge of strong discussion on the
nature of such a quantum phase and the related phase diagram \cite{discuss,liu-lieb}.

To help to clarify this issue, we performed the cluster dynamical mean field theory
calculations for the half-filled Hubbard model defined on a honeycomb lattice as,
\begin{equation}\label{eq:Hubbard_honeycomb}
\hat{H}=-t\sum_{\langle ij \rangle,\sigma} (c_{i\sigma}^\dagger c_{j\sigma} + H.c.) +
U\sum_{i} n_{i \uparrow}n_{i \downarrow},
\end{equation}
where $c^{\dagger}_{i\sigma}~(c_{i\sigma})$ is the electron creation (annihilation)
operator with spin $\sigma$ ($\uparrow$ or $\downarrow$) at lattice site $i$, $\langle ij
\rangle$ represents the summation over the nearest neighbors, $t > 0$ is the nearest
neighbor hopping integral, and $n_{i\sigma} = c_{i\sigma}^\dagger c_{i\sigma}$ with the
on-site Coulomb repulsion $U$.

\begin{figure}
\includegraphics[width=7.5cm]{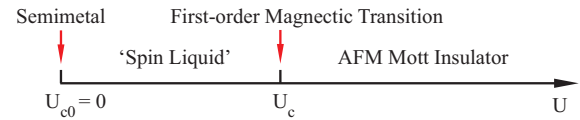}
\caption{\label{phase_diagram}(Color online) Schematic phase diagram of the one-band
Hubbard model on a honeycomb lattice at half-filling. $U_c$ is between $4.6t$ and
$4.85t$. }
\end{figure}

Our calculated results are schematically summarized in Fig. \ref{phase_diagram}. By the
calculations, we show that a disordered phase of `spin liquid' exists from $U=0$ to
$U_c$, at which it transforms into the N\'{e}el AFM insulating phase via a first-order
quantum phase transition.

The dynamical mean field theory (DMFT) maps a quantum lattice model onto a single lattice
site, a quantum impurity, dynamically coupled to a self-consistently determined bath of
free electrons that represents the rest of the lattice \cite{Georges.RMP.96}. Thus the
DMFT fully considers local quantum dynamical fluctuations, much beyond conventional
mean-field methods. The DMFT has substantially improved our understanding on the
nonperturbative properties of correlated electron systems, particularly the Mott
transition. The cluster dynamical mean field theory (CDMFT) is a natural extension of the
DMFT to include the missed short-ranged spatial correlations through a proper replacement
of a single-site impurity by a cluster of lattice sites, which is constructed to reflect
the lattice symmetry and local lattice structure features
\cite{jarrell,cell,Maier.RMP.05}. The CDMFT has been successfully applied to study a
variety of ordered phases, and opens an avenue to directly study quantum phase
transitions.

Here it should be addressed that for the (C)DMFT the thermodynamic limit is naturally
taken from the outset through a self-consistent procedure
\cite{Georges.RMP.96,Maier.RMP.05}. As a nonperturbative approach to treat many-body
correlation effects, the (C)DMFT works well in the whole coupling regime and becomes
exact in the two contrary limits of both noninteracting and infinite-interacting cases.
For finite-size quantum Monte Carlo simulations or exact diagonalizations, in contrast,
the thermodynamic limit is extrapolated through finite-size scaling. Correspondingly, the
CDMFT allows for spontaneous symmetry breaking, while the finite-size approaches have
difficulty in finding a long-range order and related phase transition or underestimate
ordered phases. Thus the CDMFT and finite-size approaches are complementary to each
other, both of which together can give more conclusive results than they alone.

In CDMFT calculations, the target in the self-consistent procedure is to obtain a lattice
imaginary frequency local Green's function matrix $G_{ij}(i\omega_n)$ (subscripts $i$ and
$j$ being site indices of a chosen cluster) by assuming its self-energy matrix identified
as the one of the corresponding cluster-impurity Green's function matrix
$G_{ij}^{imp}(i\omega_n)$, derived from the Dyson equation. In order to study a ground
state, we apply exact diagonalization rather than quantum Monte Carlo simulation to solve
a cluster-impurity model \cite{Caffarel94,Liebsch07}. Specifically, we employed the
robust Krylov-Schur algorithm based SLEPc \cite{SLEPc} to accomplish the large-scale
sparse matrix diagonalization efficiently and stably \cite{he}. Here we particularly
emphasize that to carry out an elaborate numerical analytic continuation from an
imaginary frequency Green's function $G_{ii}(i\omega_n)$ onto a real frequency retarded
Green's function $G_{ii}(\omega+i0^{+})$ is crucial to unambiguously identify whether or
not an energy gap exists at a small on-site interaction $U$
\cite{ac,analytical,analconti}, by checking the density of states (DOS) equal to
$-\frac{1}{\pi}Im G_{ii}(\omega+i0^{+})$. In such a way, the energy gap resolution can be
achieved as high as $10^{-3}t$, which is hardly reached by other methods.

\begin{figure}
\includegraphics[width=7.0cm]{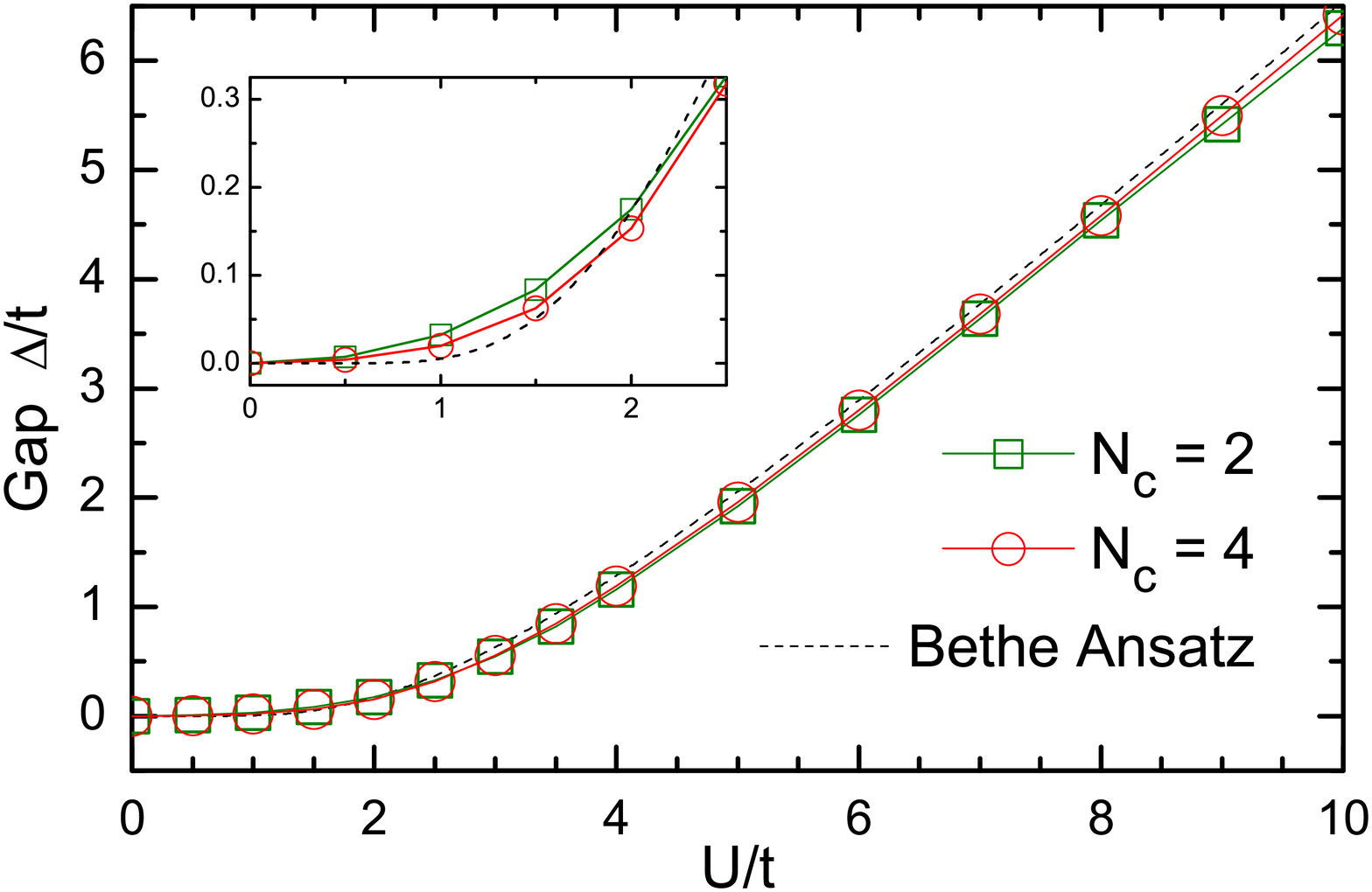}
\caption{\label{1d} (Color online) Energy gap $\Delta$, namely single-particle spectral
gap, as a function of the on-site interaction $U$ in the one-dimensional half-filled
Hubbard model, calculated with the two-site ($N_c$=$2$) and four-site ($N_c$=$4$)
clusters, respectively. The dashed curve denotes the exact one from the Bethe ansatz
solution. The inset zooms in the small-$U$ dependence. }
\end{figure}

We first performed the CDMFT calculations for one-dimensional (1D) Hubbard model at
half-filling with a cluster-impurity model respectively containing two and four lattice
impurity sites, which serves as a benchmark for the further calculations. The calculated
results are reported in Fig. \ref{1d}, in quantitative comparison with the Bethe Ansatz
exact solution \cite{1dhubbard}. As we see, the two-site CDMFT result has been already in
excellent agreement with the exact solution. Especially, by using the numerical analytic
continuation, we can unambiguously identify a finite energy gap immediately develops once
$U$ is nonzero. Thus the short-ranged spatial correlations play a dominant role in local
spectral functions and properties even though the spatial correlations have long-ranged
power-law behavior in 1D Hubbard model \cite{cor}. It is also well-known that the quantum
fluctuations are much stronger in one-dimension than in higher dimensions. Hence it is
highly expected that the higher dimensional CDMFT results are more reliable and
encouraging than the one-dimensional ones.

\begin{figure}
\includegraphics[width=2.6cm]{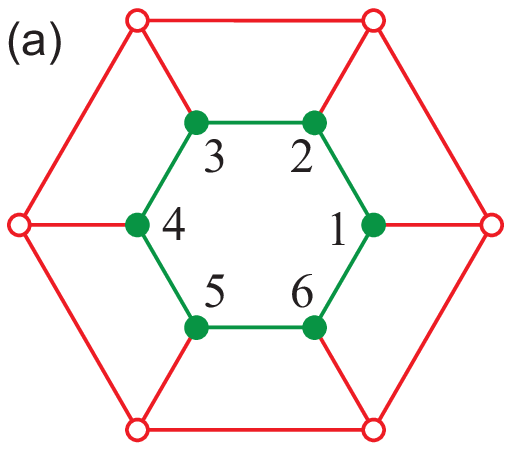}
\includegraphics[width=5.1cm]{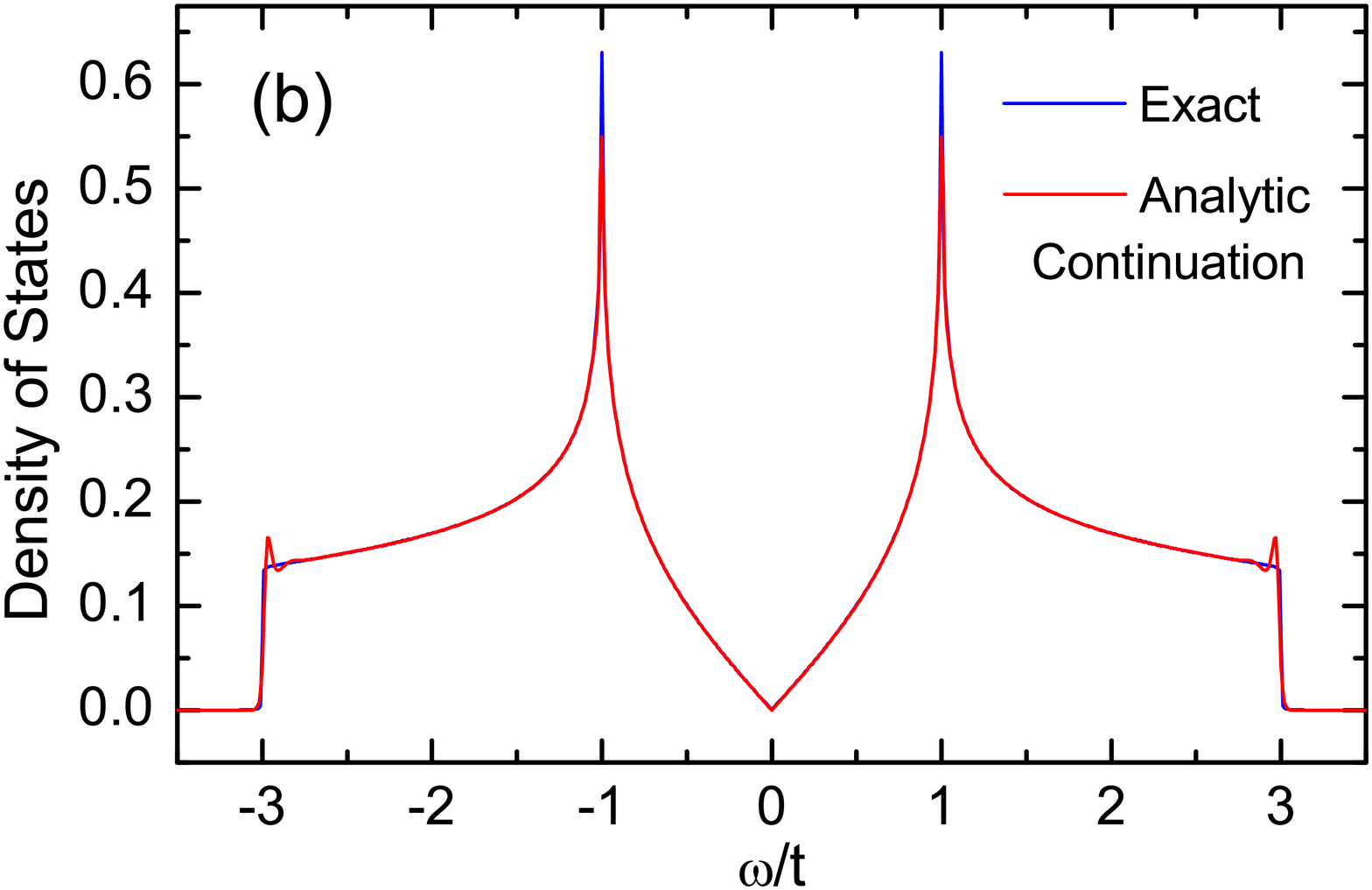}
\caption{\label{bath_levels}(Color online) (a) Cluster-impurity model configuration for a
honeycomb lattice. Filled circles denote the impurity sites. Unfilled circles denote the
bath levels. Links represent the hopping paths. (b) Density of states of the model when
$U = 0$. The blue curve is the exact one. The red one was obtained by the numerical
analytic continuation. }
\end{figure}

Figure \ref{bath_levels}(a) schematically shows the cluster-impurity model constructed
for a honeycomb lattice. Such a model reflects sixfold rotational symmetry and an
impurity site with a lattice coordination number of 3, which are the essential features
of a honeycomb lattice. We also add a direct link between each pair of the nearest bath
levels so that we can include the propagation of an electron from one impurity site
through the outside of the cluster (the bath) to any other impurity site in the
calculations.

When the on-site interaction $U=0$, the half-filled Hubbard model on a honeycomb lattice
reduces to a set of free Dirac fermions with a linear DOS around the Fermi energy. As
shown in Fig. \ref{bath_levels}(b), the numerical analytic continuation can well repeat
the exact DOS. Particularly, around the Fermi energy both are the strictly same even
though the DOS is not differentiable at the Fermi energy.

\begin{figure}
\includegraphics[width=7.0cm]{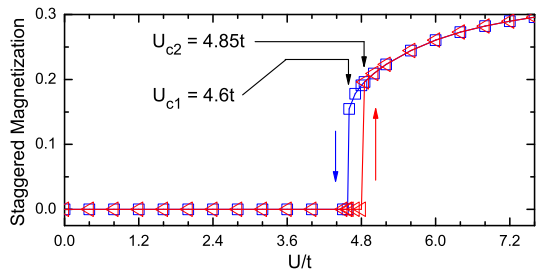}
\caption{\label{AFM} (Color online) Calculated magnetization as a function of the on-site
Coulomb interaction $U$ on a honeycomb lattice. }
\end{figure}

In this study, we define the magnetization $m$=$\langle (n_{\uparrow}-n_{\downarrow})/2
\rangle $ on a site. Being bipartite, a honeycomb lattice can be divided into two
sublattices $A$ and $B$. If a N\'{e}el AFM state appears, $m$ will alternatively take
positive and negative along with sublattices $A$ and $B$, namely being staggered
magnetization. As we see from Fig. \ref{AFM}, when $U<U_{c2}=4.85t$, a paramagnetic
solution of $m=0$ is stable, namely no spin polarization on each site, but over $U_{c2}$
this solution is no longer stable and $|m|$ abruptly jumps over 0.2. On the other hand,
when $U>U_{c1}=4.6t$ a staggered magnetization solution with $|m|>0.16$ is stable, namely
a N\'{e}el AFM phase takes over, but below $U_{c1}$ $m$ immediately plummets to zero.
Between $U_{c1}$ and $U_{c2}$, these two solutions coexist. Such a hysteresis behavior
indicates this magnetic transition is a first-order quantum phase transition.

\begin{figure}
\includegraphics[width=7.0cm]{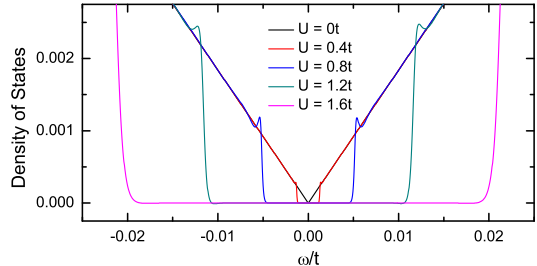}
\caption{\label{doszoomin} (Color online) Calculated density of states for several small
values of the on-site Coulomb interaction $U$ on a honeycomb lattice, zooming in around
the Fermi energy $\omega=0$. }
\end{figure}

\begin{figure}
\includegraphics[width=7.0cm]{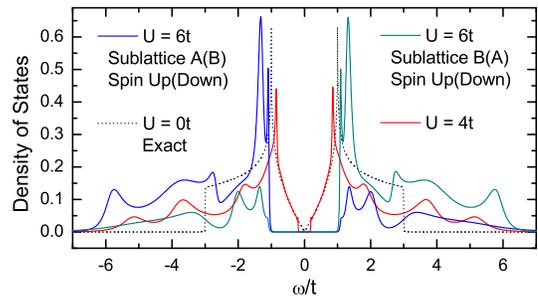}
\caption{\label{dos} (Color online) Calculated density of states for $U = 4t$ (red lines)
and $U=6t$ (blue and green lines) on a honeycomb lattice, respectively. The dotted curve
is the exact DOS at $U=0$ for a reference and the Fermi energy sets to zero. }
\end{figure}

We calculated the DOS for a small on-site interaction $U$ with extreme caution through
elaborate numerical analytic continuation \cite{analytical}. The calculated DOS are then
plotted in Fig. \ref{doszoomin}. Similar to the case of 1D Hubbard model, what we find is
that there is also a definite energy gap opening at the Fermi energy once $U$ is nonzero.
In comparison with the case of $U=0$ by checking the enclosed area, it is further shown
that the corresponding states nearby the Fermi energy are clearly moved away from the gap
rather than pushed to the two sides of the gap. To be specific, for $U=0.4t,~0.8t,$ and
$1.2t$, the energy gap is found as small as $\Delta=2.5\times 10^{-3}t, 1.0\times
10^{-2}t,$ and $0.023t$, respectively. For a rather large $U$, a relatively large energy
gap opens with a substantial portion of states moved away from the Fermi energy into
below $-3t$ and above $3t$, corresponding to the Hubbard band states. For a $U$ further
larger than $U_{c2}$, the system transforms into the N\'{e}el AFM phase. The on-site spin
degeneracy is then lifted. The spin-up (spin-down) resolved DOS at $A$ sublattice is the
same as the spin-down (spin-up) resolved DOS at $B$ sublattice, as shown in Fig.
\ref{dos}.

\begin{figure}
\includegraphics[width=7.0cm]{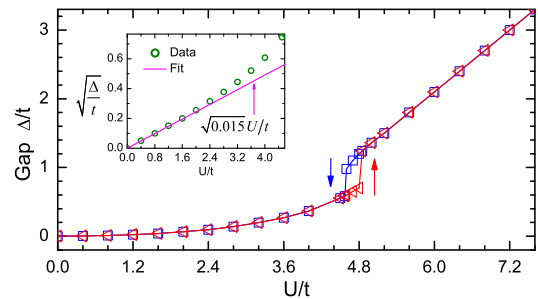}
\caption{\label{gaps} (Color online) Calculated energy gap $\Delta$ as a function of the
on-site Coulomb interaction $U$ on a honeycomb lattice. The inset zooms in the small-$U$
dependence. Note that the energy gap is in linear $U$-dependence after the transition. }
\end{figure}

Figure \ref{gaps} shows the energy gap $\Delta$ as a function of the on-site interaction
$U$, extracted from the calculated DOS. For $U<1.6t$, the energy gap increases very
slowly with $U$, and the function can be represented as $\Delta/t = 0.015(U/t)^2$. After
$1.6t$, the energy gap increasing becomes fast with $U$ increasing. Between $U_{c1}$ and
$U_{c2}$, the $U$-dependence of the energy gap shows a hysteresis behavior with a sudden
change of $\sim$$0.45t$, corresponding to the first-order quantum phase transition. Thus
a nonzero $U$ definitely induces an energy gap and makes the system be in an insulating
phase.

We also calculated the equal time spin-spin correlation functions $\langle s_{i}^{z}
s_{j}^{z} \rangle$ and occupancy-occupancy correlation functions $\langle n_{i} n_{j}
\rangle$, with $i,~j$ being a pair of the nearest sites, next nearest sites, and next
next nearest sites, respectively. We find that all these correlation functions keep the
sixfold rotational symmetry as the honeycomb lattice does before the magnetic transition.
This excludes spin valence bond structures, spin anisotropic structures, and charge
density waves. We can also exclude another possibility to break the translation symmetry
that every six sites form a benzene-ring-like plaquette to further form a
plaquette-singlet valence bond solid pattern in a honeycomb lattice. If such a pattern
exists, the states nearby the Fermi energy will be pushed to the two sides of the gap
rather than moved away. Thus this nonmagnetic insulating phase from $U=0$ to $U_c$
(between $4.6t$ and $4.85t$) can be classified as a `spin liquid' phase in the sense that
it is tuned on by the on-site interaction $U$.

\begin{figure}
\includegraphics[width=7.0cm]{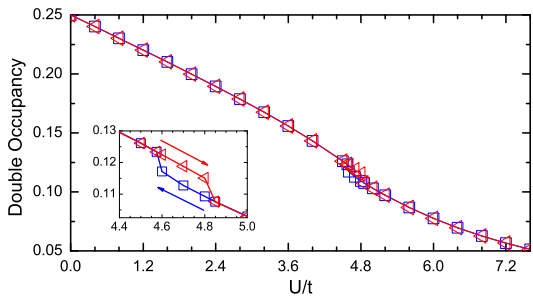}
\caption{\label{docc} (Color online) Calculated double occupancy $D$ as a function of the
on-site Coulomb interaction $U$ on a honeycomb lattice. The inset zooms in the hysteresis
loop. }
\end{figure}

To understand the underlying physics, we examine the double occupancy, defined as
$D=\langle n_{\uparrow}n_{\downarrow}\rangle$ on a site. The ground state energy per site
$E_g=\langle \hat{H} \rangle/N$ of Hamiltonian (\ref{eq:Hubbard_honeycomb}) is a function
of the on-site interaction $U$. Its derivative is nothing but the double occupancy,
namely $\partial {E_g}/\partial{U} =\langle n_{\uparrow}n_{\downarrow} \rangle$. Thus the
double occupancy $D$ directly describes a quantum phase transition tuned by $U$. In Fig.
\ref{docc}, the $U$-dependence of the $D$ likewise shows a hysteresis behavior between
$U_{c1}$ and $U_{c2}$. This means the energy level crossing in the ground state through
the magnetic transition as $U$ increasing, being a characteristic of a first-order
quantum phase transition \cite{book1}.

The on-site interaction $U$ tunes or controls the Hamiltonian (1) through the double
occupancy $D$. Moreover, the localization degree of an electron, as well as the local
correlation effect, can be quantitatively described by the double occupancy. At
half-filling, $D$ is between 0.25 and 0, corresponding respectively to full
delocalization and complete localization. In addition, the magnetic moment $m_z$ is
obtained by $\langle m_z^2 \rangle=\langle (2S^z)^2 \rangle=1-2D$.

It is commonly thought that the Mott transition is driven by the strong local correlation
effect due to the on-site interaction $U$, which is marked by a vanishing or very small
double occupancy with a large local moment on a site \cite{Georges.RMP.96,rice}. In
contrast, for the Hubbard model on a honeycomb lattice, a small $U$ can immediately
induce a small energy gap opening to tune the system into an insulating phase with a
large double occupancy namely large charge fluctuation, as shown in Fig. \ref{docc}. The
calculations show that the small-$U$ induced energy gap is a consequence of the interplay
between the zero-DOS at the Fermi energy (Dirac Cone band) and local charge correlation,
not a conventional correlation-driven Mott insulating gap. On the other hand, the
correlation effect will become dominating nearby the magnetic transition. Actually Fig.
\ref{gaps} has shown that the energy gap becomes a linear function of $U$ after the
transition, which is the canonical behavior of a correlation-driven Mott insulator. Thus
the `spin liquid' states nearby $U=0$ are of charge fluctuation dominating while those
nearby $U_c$ are of charge correlation dominating, and corresponding to the ones found by
the quantum Monte Carlo simulations reported in Ref. \onlinecite{Meng10}. Nevertheless
our calculations show that a smooth crossover connects these two contrary parts.
Meanwhile it is also indicated that the charge fluctuation plays a substantial role on
keeping the `spin liquid' phase against the emergence of an AFM order.

In summary, we have performed the cluster dynamical mean field theory calculations,
allowing for the spontaneous symmetry breaking, to study the ground state of the
half-filled Hubbard model on a honeycomb lattice. We find that a `spin liquid' exists
from $U=0$ to about $4.6t$, in which the system takes a smooth crossover correspondingly
from the charge fluctuation dominating phase into the charge correlation dominating
phase, then it further transforms into the N\'{e}el AFM Mott insulating phase via a
first-order quantum phase transition.


\begin{acknowledgments}

We would like to thank Ning-Hua Tong for very helpful discussions. ZYL sincerely thanks
the hospitality of International Center of Quantum Materials of Peking University, where
this manuscript was finalized. This work is partially supported by National Natural
Science Foundation of China and by National Program for Basic Research of MOST
(2011CBA00112), China.

\end{acknowledgments}


\begin{thebibliography}{}
\bibitem{he} We rewrote the code from scratch in a fully parallel computimg structure to
exploit the parallel framework of SLEPc and meanwhile be able to incorporate a variety of
cluster-impurity model configurations.
\bibitem{analytical} We adopted the Pad\'{e} approximant algorithm for numerical
analytic continuation \cite{analconti}. An effective temperature of $10^{-5}t$ is set for
a small U, and $5\times 10^{-3}t$ for a large U.
\end{thebibliography}
\end{document}